\documentclass[12pt]{article}

\usepackage{geometry}
\usepackage[parfill]{parskip}

\usepackage{amsmath, amssymb, amsthm}

\usepackage[ruled,vlined]{algorithm2e}
\usepackage{listings}
\usepackage{graphicx}
\usepackage{url}
\usepackage{xcolor}

\usepackage{caption}
\usepackage{subcaption}

\title{An approach for benchmarking the numerical solutions of stochastic compartmental models}
\author{Alison C. Hale*, Christopher P. Jewell\\ *a.c.hale@lancaster.ac.uk\\ *Department of Mathematics and Statistics, Lancaster University, Lancaster, UK.}

\begin{document}

\maketitle

\bigskip
\bigskip

\begin{center}
\textbf{Abstract}
\end{center}

\leftskip1.5cm
\rightskip1.5cm

    An approach is introduced for comparing the estimated states of stochastic compartmental models for an epidemic or biological process with analytically obtained solutions from the corresponding system of ordinary differential equations (ODEs).
    Positive integer valued samples from a stochastic model are generated numerically at discrete time intervals using either the chain Binomial or Gillespie algorithm.
    The simulated distribution of realisations is compared with an exact solution obtained analytically from the ODE model.
    Using this novel methodology this work demonstrates it is feasible to check that the realisations from the stochastic compartmental model adhere to the ODE model they represent.
    There is no requirement for the model to be in any particular state or limit.
    These techniques are developed using the stochastic compartmental model for a susceptible-infected-recovered (SIR) epidemic process.
    The Lotka–Volterra model is then used as an example of the generality of the principles developed here.
    This approach presents a way of testing/benchmarking the numerical solutions of stochastic compartmental models, e.g. using unit tests, to check that the computer code along with its corresponding algorithm adheres to the underlying ODE model.

\leftskip0cm
\rightskip0cm

\bigskip
\bigskip
\bigskip

\clearpage
\newpage

\section{Introduction}

    It is of vital importance when developing computer software to have testing procedures that ensure it functions as required.
    Testing aims to provide objective information about the quality of the software and its risks of failure.
    In general testing may include checking functionality given valid and invalid inputs, potential concurrency issues, security vulnerabilities and so on.
    Consequently testing during the software development and test cycles seeks to improve functionality and fix bugs.
    Moreover tests from the previous and current development cycles may be of use in helping to reduce the risk of malfunctions in future development cycles.
    However despite extensive literature on software testing, e.g. \cite{Tarlinder2016, Black2013, Myers2012}, it is too often given inadequate attention leading to avoidable errors \cite{Merali2010}.
    
    Testing frameworks are not only needed for assessing the functionality of software packages but are also vital for benchmarking numerical software.
    The latter requires an adequate amount of dynamic testing \cite{Fairley1978}.
    Floating-point issues aside it may be a significant challenge to find a suitable mathematical regime under which to perform sufficiently rigorous tests to verify the numerical solutions of mathematical models.
    As such benchmarking numerical solutions is often applied to very specific models motivated by a particular application for example: Proposal for numerical benchmarking of fluid-structure interaction between an elastic object and laminar incompressible flow \cite{Turek2006}, Benchmarking of numerical integration methods for ODE models of biological systems \cite{Stadter2020}, Benchmarking and developing numerical finite element models of volcanic deformation \cite{Hickey2014} and Benchmarking five numerical simulation techniques for computing resonance wavelengths and quality factors in photonic crystal membrane line defect cavities \cite{lasson2018}.
    This is by no means an exhaustive list since in general a given class of equations or models has a corresponding set of solutions therefore with regard to benchmarking numerical solutions each class will need to be considered on an individual basis.
    
    This paper introduces a procedure for numerically testing the solutions from systems of ordinary differential equations (ODEs) which are solved numerically using stochastic methods.
    This is of particular interest when the system of ODEs cannot be solved analytically.
    Although the basic procedure proposed in this paper may be easily adapted to systems of ODEs which are solved numerically by a purely deterministic method such as the Euler method, the primary interest here is the arguably harder problem of testing the validity of numerical solutions produced using stochastic methods.
    The purpose of this paper is to introduce a novel testing/benchmarking procedure for stochastic compartmental models.
    Given such a model the chain Binomial algorithm \cite{Abbey1952, Fine1977} and/or Gillespie algorithm \cite{Gillespie1976, Gillespie1977} are used to simulate realisations of the solution at discrete time points.
    The distribution of realisations is compared using statistical techniques with an exact solution derived analytically from the underlying system of ODEs.
    The entire time evolution of the system of ODEs is included in this novel testing procedure, hence there is no requirement for the model to be in any particular state e.g. a thermodynamic limit or a late time steady state.
    Additionally this testing procedure does not rely upon setting a random seed or using a prespecified system architecture.
    The new methodology presented here therefore checks that the simulated realisations adhere to the underlying system of ODEs. Importantly this allows the solutions produced by computer code to be benchmarked and in a more formal testing setting this procedure could form the basis of a unit test.

    To demonstrate this novel testing approach the stochastic compartmental model for a susceptible-infected-recovered (SIR) epidemic process is used as an exemplar: this class of models is well documented for example see textbook \cite{Brauer2019}.
    The Lotka–Volterra, predator-prey, model \cite{Lotka1925, Volterra1926} is used to demonstrate both the generality of the testing approached and highlight subtle differences in its application to both models e.g. differences in the form of the numerical solutions.
    The extent to which the approach presented in this paper can be applied to arbitrary compartmental models is explored in the Discussion section.

\section{Outline of testing approach}
\label{section:outlineMeth}

    The following benchmarking procedure is proposed given the realisations (solution) computed numerically from a stochastic compartmental model:
    \begin{enumerate}
        \item For a given set of parameters simulate realisations from the epidemic or biological process using for example a chain Binomial or Gillespie algorithm.
        \item Fully or partially integrate analytically the system of ODEs describing the compartmental model. Rearrange the solution into an \emph{expression} in terms of one, or a combination of, parameters (i.e. parameters from the system of ODEs).
        \item Compute estimates of the parameter (or combination of parameters) by substituting the aforementioned realisations (from part 1.) into the \emph{expression} (from part 2.).
        \item Compute the mode of the distribution of parameter estimates (from part 3.) and check it equals, within an acceptable degree of uncertainty, the actual parameter value.
        \item Where viable repeat this procedure as required for other parameters, or combination of parameters, in the system of ODEs.
    \end{enumerate}
    
    In general, given a the system of ODEs for a stochastic compartmental model, any applicable method can be used to fully or partially integrate analytically.
    In this paper it is convenient to use the separation of variables technique.

\section{SIR model}
\subsection{System of ODEs}
    The system of ODEs for the continuous SIR model \cite{Brauer2019, Kermack1927, Hethcote2000} with $S(t)$ susceptible, $I(t)$ infected and $R(t)$ recovered individuals at time $t$ is
    \begin{equation}
    \label{eq:ODE1} 
    \frac{dS(t)}{dt} =  \frac{-\beta I(t)}{\mathcal{N}} S(t),
    \end{equation}
    \begin{equation}
    \label{eq:ODE2}
    \frac{dI(t)}{dt} =  \frac{\beta I(t)}{\mathcal{N}} S(t) - \gamma I(t),
    \end{equation}
    \begin{equation}
    \label{eq:ODE3}
    \frac{dR(t)}{dt} =\gamma I(t).
    \end{equation}
    
    Constant $\mathcal{N}$ represents the total number of individuals in the population while real constants $\beta>0$ and $\gamma>0$ determine the rate a which individuals move between states $S \rightarrow I$ and $I \rightarrow R$ respectively.
    Clearly $dS(t)/dt+dI(t)/dt+dR(t)/dt=0$ from which it follows that $\mathcal{N}$ is conserved i.e. $S(t)+I(t)+R(t)=\mathcal{N}$.
    Consequently this SIR model represents a closed system in other words individuals do not flow in or out of the system via birth, death, migration, or any other means. The number of individuals flowing between states $S \rightarrow I$ and $I \rightarrow R$ can respectively be expressed in terms of the number of events $N_{SI}(t)$ and $N_{IR}(t)$ as follows
    \begin{equation}
    \label{eq:ODEeventsSI}
    \frac{dN_{SI}(t)}{dt} =  \frac{-\beta I(t)}{\mathcal{N}} S(t),
    \end{equation}
    \begin{equation}
    \label{eq:ODEeventsIR}
    \frac{dN_{IR}(t)}{dt} =  \gamma I(t).
    \end{equation}

    \bigskip

    Given the chain rule $dS/dt=dS/dR \,\, dR/dt\,$ it follows from Equations \ref{eq:ODE1} and \ref{eq:ODE3} that
    \begin{equation}
    \frac{dS(t)}{dR(t)} =  \frac{-\beta S(t)}{\gamma \mathcal{N}}.
    \end{equation}
    Separation of variables may be used to solve this differential equation hence
    \begin{equation}
    \int_{S(0)}^{S(t)} \frac{1}{S} \,dS = \frac{-\beta}{\gamma \mathcal{N}} \int_{R(0)}^{R(t)} \,dR.
    \end{equation}
    The corresponding solution to this integral equation is 
    \begin{equation}
    \label{eq:transc}
    \ln \frac{S(t)}{S(0)} =  \frac{\beta}{\gamma \mathcal{N}} (R(0) - R(t)).
    \end{equation}
    This solution is a transcendental equation which cannot be solved analytically.
    However rearranging Equation \ref{eq:transc} in terms of $R(\infty)$ and noting the number of individuals is conserved leads to
    \begin{equation}
    \label{eq:transcExp}
    R(\infty) = \mathcal{N}-S(0) \exp\left({\frac{\beta}{\gamma \mathcal{N}} R(0)}\right) \exp\left({\frac{-\beta}{\gamma \mathcal{N}} R(\infty)}\right)
    \end{equation}
    where $I(\infty)=0$.
    Given the Lambert $W_k$ function, the solution to $x=a+be^{cx}$ is $x=a-c^{-1}W_k(-bce^{ac})$ where $a$, $b$, and $c$ are complex constants, $b$ and $c$ are not equal to zero, and $k$ is an integer. Therefore the solution to Equation \ref{eq:transcExp} in terms of the Lambert $W_k$ function is 
    \begin{equation}
    \label{eq:lambertW0}
    R(\infty) = \mathcal{N} - \left(\frac{-\beta}{\gamma\mathcal{N}}\right)^{-1} W_k\left( \frac{-\beta}{\gamma\mathcal{N}} S(0) \exp \left( \frac{-\beta}{\gamma\mathcal{N}}(\mathcal{N}-R(0)) \right) \right)  
    \end{equation}
    With $k=0$ (principle branch) then $W_0(\cdot)$ has a single real value provided $R(t) \geqslant 0 \,\,\, \forall \, t$.
    
    A computer algorithm which approximates Equations \ref{eq:ODE1}~to~\ref{eq:ODE3} using a stochastic compartmental model produces a distribution of realisations which are expected to adhere, at least in a stochastic sense, to the underlying system of ODEs.
    Due to the inherent randomness of such a method none of the Equations \ref{eq:transc}, \ref{eq:transcExp} or \ref{eq:lambertW0} are in a convenient form to check that the realisations, at each time step, follow the underlying system of ODEs.
    However rearranging Equation \ref{eq:transc} in terms of the reproduction number $\beta/\gamma$ yields
    \begin{equation}
    \label{eq:boverg}
    \frac{\beta}{\gamma} = \frac{\mathcal{N}}{R(0) - R(t)}
    \ln \frac{S(t)}{S(0)}.
    \end{equation}
    All time dependent factors are on the right side hence the left side is time invariant.
    It is therefore reasonable to expect that any computer algorithm which approximates Equations \ref{eq:ODE1}~to~\ref{eq:ODE3} using a stochastic compartmental model will, given Equation \ref{eq:boverg}, produce reproduction number estimates at every time step provided $S(t) \neq S(0)$, $R(0) \neq R(t)$, $S(t) \neq 0$, and/or $S(0) \neq 0$.
    In practice, this will give rise to a distribution of reproduction number estimates where the mode equals the parameter ratio $\beta/\gamma$.
    Note it is not the objective here to seek solutions to this system of ODEs.

\subsection{Stochastic compartmental model}

    Numerical solutions to Equations \ref{eq:ODE1}~to~\ref{eq:ODE3} are sought for the number of individuals in each state. For the SIR stochastic compartmental model let the state variables be denoted $\tilde{S}$, $\tilde{I}$ and $\tilde{R}$.
    The events, the number of individuals moving between states during a given time interval, will be denoted $\tilde{N}_{SI}$ and $\tilde{N}_{IR}$ for the state transitions $S \rightarrow I$ and $I \rightarrow R$ respectively.
    Let $\Delta$ denote an integer increment in a process over a finite time interval $[t,t+\delta t)$.
    For example the incremental change in the number of $S \rightarrow I$  events is $\Delta \tilde{N}_{SI}(t) = \tilde{N}_{SI}(t+\delta t) - \tilde{N}_{SI}(t)$.
    The discrete analogue of Equations \ref{eq:ODE1}~to~\ref{eq:ODE3} is
    \begin{equation}
    \Delta \tilde{S} = - \Delta \tilde{N}_{SI}(t),
    \end{equation}
    \begin{equation}
    \Delta \tilde{I} =   \Delta \tilde{N}_{SI}(t) - \Delta \tilde{N}_{IR}(t),
    \end{equation}
    \begin{equation}
    \Delta \tilde{R} =   \Delta \tilde{N}_{IR}(t).
    \end{equation}
    With reference to Equations \ref{eq:ODEeventsSI} and \ref{eq:ODEeventsIR} let small increments in the number of events be denoted as $\delta \tilde{N}_{SI} = \tilde{\mu}_{SI} \tilde{S}(t) \delta t$ and $\delta \tilde{N}_{IR} = \tilde{\mu}_{IR} \tilde{I}(t) \delta t$ where for convenience  $\tilde{\mu}_{SI}(t) = \beta I(t)/\mathcal{N}$ and $\tilde{\mu}_{IR}=\gamma$. 
    
    In the following two commonly used algorithms, chain Binomial and Gillespie, are used in the context of the SIR model.  Given there are various versions of these algorithms they will be explicitly defined below for clarity.

    \subsubsection{chain Binomial algorithm}
    There are several stochastic Euler schemes based on the chain Binomial method \cite{Abbey1952, Fine1977} which could be used to determine the number of events $\Delta \tilde{N}_{SI}(t)$ and $\Delta \tilde{N}_{IR}(t)$.
    For example a Poisson distribution where at each time increment $\Delta \tilde{N}_{SI}(t) \sim \mathrm{Poi}(\tilde{\mu}_{SI}(t) \tilde{S(t)} \delta t)$ and $\Delta \tilde{N}_{IR}(t) \sim \mathrm{Poi}(\tilde{\mu}_{IR} \tilde{I(t)} \delta t)$.
    However here the focus is on the Binomial distribution where the realisations of events are given by
    \begin{equation}
    \Delta \tilde{N}_{SI}(t) \sim \mathrm{Bin}(\tilde{S}(t), \,\, 1-\exp(-\tilde{\mu}_{SI}(t) \delta t)),
    \end{equation}
    \begin{equation}
    \Delta \tilde{N}_{IR}(t) \sim \mathrm{Bin}(\tilde{I}(t), \,\, 1- \exp(-\tilde{\mu}_{IR} \delta t)).
    \end{equation}
    Hence the probability, conditional on all states, of one infection in time period $\delta t$ is given by the cumulative distribution function of the Exponential distribution.
    It follows that the number of individuals in each state at time $t$ is
    \begin{equation}
    \label{eq:solA}
    \tilde{S}(t) =  \tilde{S}(0) - \sum_{\tau=1}^{t} \Delta \tilde{N}_{SI}(\tau \delta t),
    \end{equation}
    \begin{equation}
    \label{eq:solB}
    \tilde{I}(t) = \tilde{I}(0) + \sum_{\tau=1}^{t} \left(\Delta \tilde{N}_{SI}(\tau \delta t) - \Delta \tilde{N}_{IR}(\tau \delta t) \right),
    \end{equation}
    \begin{equation}
    \label{eq:solC}
    \tilde{R}(t) = \tilde{R}(0) + \sum_{\tau=1}^{t} \Delta \tilde{N}_{IR}(\tau \delta t).
    \end{equation}

    The chain Binomial algorithm uses the Euler–Maruyama approximation to compute the states of this system at each time step: see Algorithm~\ref{al:cb}.
    In this algorithm the number of individuals arriving in each state at a given time step depends on the number in the corresponding state at the previous time step.
    Consequently the realisations from each state form a first-order Markov process.
    
    \bigskip
    
    \begin{algorithm}[H]
    \SetAlgoLined
    $\delta t>0$, $t=0,1,...,T$\\
    $\beta>0$, $\gamma>0$\\
    $S_0>0$, $I_0>0$, $R_0\geqslant0$\\
    $\mathcal{N}=S_0+I_0+R_0$\\
    \While{$(t<T$ $\mathrm{and}$ $I_t \neq0)$}{
    
    Draw:\\
    \hspace{1em}$\Delta N_{SI} \sim \mathrm{Bin}(S_t, \, 1-\exp(-\beta I_t \mathcal{N}^{-1} \delta t$)\\
    \hspace{1em}$\Delta N_{IR} \sim \mathrm{Bin}(I_t, \, 1-\exp(-\gamma \delta t$)\\
    Let:\\
    \hspace{1em}$S_{t+1} = S_t - \Delta N_{SI} $\\
    \hspace{1em}$I_{t+1} = I_t + \Delta N_{SI} - \Delta N_{IR} $\\
    \hspace{1em}$R_{t+1} = R_t + \Delta N_{IR}$\\
    $t = t + 1$
    }
    \caption{chain Binomial algorithm for SIR model}
    \label{al:cb}
    \end{algorithm}

    \bigskip

    \subsubsection{Gillespie algorithm}
    The Gillespie scheme \cite{ Gillespie1976, Gillespie1977, Doob1942, Doob1945} is a stochastic Euler scheme which differs from the chain Binomial scheme in that time increment length varies stochastically, and the transition size per time increment is fixed such that $\Delta \tilde{N}_{SI}(t) = \pm 1$, $\Delta \tilde{N}_{IR}(t) = \pm 1$ and $\Delta \tilde{N}_{IR} = \pm 1$.
    Given this scheme for an SIR model then:
    \begin{itemize}
        \item $S \rightarrow I$: time to next infection, conditional on $\tilde{S}(t)$ and $\tilde{I}(t)$, is drawn from $\mathrm{Exp}(\tilde{\mu}_{SI}(t) \tilde{S}(t))$.
        \item $I \rightarrow R$: time to next removal, conditional on $\tilde{I}(t)$, is drawn from $\mathrm{Exp}(\tilde{\mu}_{IR}  \tilde{I}(t))$.
    \end{itemize}
    With $\tilde{\mu}_{SIR}(t) = \tilde{\mu}_{SI}(t) \tilde{S}(t) + \tilde{\mu}_{IR}(t) \tilde{I}(t)$ it follows that the time to the next event, conditional on $\tilde{S}(t)$ and $\tilde{I}(t)$, is drawn from $\mathrm{Exp}(\tilde{\mu}_{SIR}(t))$. Therefore the probabilities of infection ($S \rightarrow I$) and removal ($I \rightarrow R$) are:
     \begin{itemize}
        \item $\mathrm{Pr}(\mathrm{infection}|\tilde{S}(t), \tilde{I}(t), \tilde{R}(t) ) = \tilde{\mu}_{SI}(t) \tilde{S}(t) \, / \, \tilde{\mu}_{SIR}(t)$.
        \item $\mathrm{Pr}(\mathrm{removal}|\tilde{S}(t), \tilde{I}(t), \tilde{R}(t) ) = 1 - \mathrm{Pr}(\mathrm{infection}|\tilde{S}(t), \tilde{I}(t), \tilde{R}(t) ) = \tilde{\mu}_{IR}(t) \tilde{I}(t) \, / \, \tilde{\mu}_{SIR}(t)$.
    \end{itemize}    

    A Gillespie algorithm for the SIR model is given in Algorithm \ref{al:gl}.
    
    \bigskip
    
    \begin{algorithm}[H]
    \SetAlgoLined
    $t=0,1,...,T$\\
    $\beta>0$, $\gamma>0$\\
    $S_0>0$, $I_0>0$, $R_0\geqslant0$\\
    $\mathcal{N}=S_0+I_0+R_0$\\
    \While{$(t<T$ $\mathrm{and}$ $I_t \neq0)$}{
    Draw:\\
    \hspace{1em}$\tau \sim \mathrm{Exp}(\mu_{SIR})$\\
    Choose index $i$ from list $[.\,,\,.]$:\\
    \hspace{1em}$i \sim \mathrm{Discrete}([\mu_{SI} S_t \, / \, \mu_{SIR}, \,\, \mu_{IR} I_t \, / \, \mu_{SIR}])$\\
    \If{$i=0$}{
        $S_{t+1} = S_{t} - 1$\\
        $I_{t+1} = I_{t} + 1$\\
    }\Else{
        $I_{t+1} = I_{t} - 1$\\
        $R_{t+1} = R_{t} + 1$\\
    }
    $t = t + \tau$\\
    }
    \caption{Gillespie algorithm for SIR model}
    \label{al:gl}
    \end{algorithm}

    \bigskip
    \bigskip

    Note that similarly to the chain Binomial algorithm the number of individuals in each state at a given time is given by Equations~\ref{eq:solA}~to~\ref{eq:solC}.

\subsection{Verifying solutions and testing software}
    The realisations from the stochastic compartmental model are expected to be consistent with the solution obtained analytically, Equation \ref{eq:boverg}.
    Rewriting Equation \ref{eq:boverg} in terms of the numerically computed states $\tilde{S}(t)$ and $\tilde{R}(t)$, given by Equations \ref{eq:solA} and \ref{eq:solC}, leads to
    \begin{equation}
    \label{eq:tildebovergA}
    \frac{\tilde{\beta}}{\tilde{\gamma}} = \frac{\mathcal{N}}{\tilde{R}(0) - \tilde{R}(t)}
    \ln \frac{\tilde{S}(t)}{\tilde{S}(0)}  \quad \forall t.
    \end{equation}
    Estimates of the reproduction number are denoted $\tilde{\beta}/\tilde{\gamma}$.
    Given realisations of the states, e.g. computed by Algorithm~\ref{al:cb} or \ref{al:gl}, it is expected that $\beta / \gamma$ will equal the mode of the distribution of $\tilde{\beta} / \tilde{\gamma}$ estimates.
    This is checked, using Algorithm \ref{al:r0}, where multiple simulations of the epidemic process are generated from either Algorithm~\ref{al:cb} or \ref{al:gl}.
    Note in Algorithm \ref{al:r0} that $\tilde{\beta}/\tilde{\gamma}$ is denoted as $r_0$.
    
    \bigskip
    
    \begin{algorithm}[H]
    \SetAlgoLined
    $T$ total number of time steps\\
    $J$ total number of simulations\\
    \For{$j=0$ \KwTo $J$ }{
        Simulate using Algorithm~\ref{al:cb} or \ref{al:gl}:\\
        \hspace{1em}$S[j,0...T]$;
        \hspace{1em}$I[j,0...T]$;
        \hspace{1em}$R[j,0...T]$;\\
    }
    $N = S[0,0]+I[0,0]+R[0,0]$\\
    \For{$j=0$ \KwTo $J$ }{
        \For{$t=0$ \KwTo $T$ }{
            \If{($S[j,t] \neq S[j,0]$ $\mathrm{and}$ $R[j,t] \neq R[j,0]$ $\mathrm{and}$ $S(t)  \neq 0$ $\mathrm{and}$ $S(0) \neq 0$)}{
                $r_0[j,t] = N \ln(\left. S[j,t] \,/ \,S[j,0]) \, \right/  (R[j,0] - R[j,t])$
            }\Else{
                $r_0[j,t]=\mathrm{NaN}$
            }  
        }
        $\overline{r}_0[j] = \mathrm{mean}(r_0[j,0...T], \mathrm{excluding \, NaNs})$ \,\,\, \# useful for Normal distributions
    }
    Estimate the mode of the $r_0$ distribution 
    \caption{Estimate ${r}_0$ for SIR model}
    \label{al:r0}
    \end{algorithm} 

    \bigskip

    There is a tacit assumption in Algorithm~\ref{al:r0} that most of the simulated processes have a good proportion of the population flowing from state $\tilde{S}(t)$ to $\tilde{R}(t)$ during the course of the epidemic: under these conditions the mode of the distribution of $\tilde{\beta}/\tilde{\gamma}$ estimates (i.e. $r_0$ estimates) will approximately equal $\beta / \gamma$.
    If most of the epidemics die out very quickly with only one or a few infections in total then it is very likely that there will be significant discrepancy between the mode of the $\tilde{\beta}/\tilde{\gamma}$ estimates and $\beta / \gamma$ due to insufficient realisations.
    However this is not an issue as Algorithm~\ref{al:r0} is intended to be used for testing/benchmarking purposes therefore suitable model parameters can be chosen.

    \subsection{Results for chain Binomial algorithm}
\begin{figure}[htb]
        \centering
        \begin{subfigure}[t]{0.49\textwidth}
            \caption{Timeseries}
            \label{fig:Figure_1a_cb_ts}
            \vspace{-0.2cm}
            \centering
            \includegraphics[width=\linewidth]{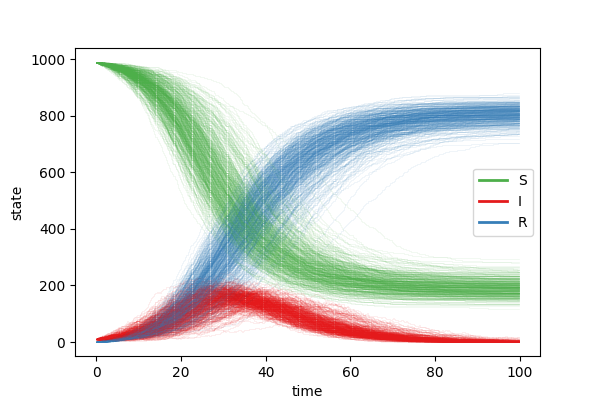} 
        \end{subfigure}

        \vspace{1.0cm}

        \begin{subfigure}[t]{0.49\textwidth}
            \caption{Histogram: mean $\tilde{\beta}/\tilde{\gamma}$}
            \label{fig:Figure_1b_cb_MeanPerSim}
            \vspace{-0.2cm}
            \centering
            \includegraphics[width=\linewidth]{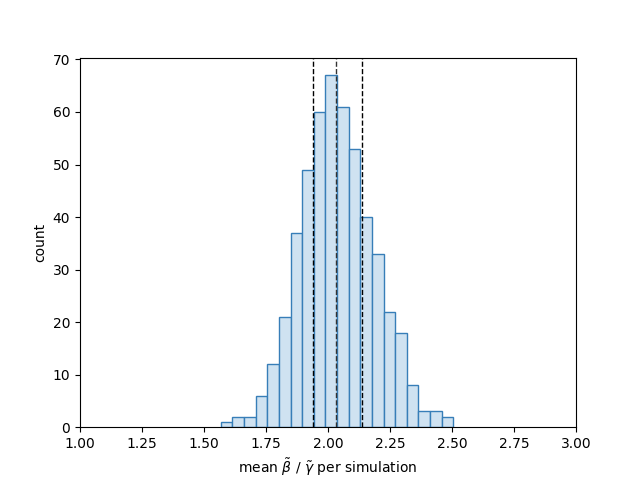} 
        \end{subfigure}
        \hfill
        \begin{subfigure}[t]{0.49\textwidth}
            \caption{Histogram: all $\tilde{\beta}/\tilde{\gamma}$}
            \label{fig:Figure_1c_cb_MeanAll}
            \vspace{-0.2cm}
            \centering
            \includegraphics[width=\linewidth]{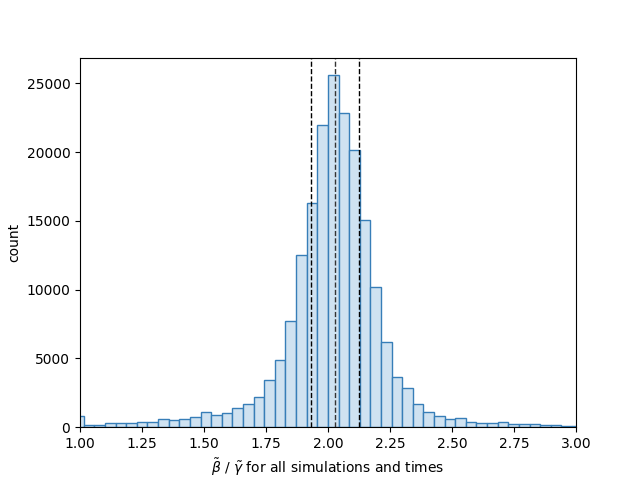} 
        \end{subfigure}

        \caption{500 simulations of the SIR epidemic process using the Chain Binomial algorithm.
        The top panel (\subref{fig:Figure_1a_cb_ts}) shows the timeseries for states $\tilde{S}(t)$, $\tilde{I}(t)$ and $\tilde{R}(t)$ for each simulated process.
        Panel (\subref{fig:Figure_1b_cb_MeanPerSim}) gives a histogram of the mean $\tilde{\beta}/\tilde{\gamma}$ from each simulation i.e. $\Bar{r}_0$.
        Lastly, panel (\subref{fig:Figure_1c_cb_MeanAll}) depicts a histogram of $\tilde{\beta}/\tilde{\gamma}$ at every time step, i.e. ${r}_0$, across all simulations. Note that ${r}_0$ and $\Bar{r}_0$ are defined in  Algorithm~\ref{al:cb}.}
    \end{figure}

    The python package \texttt{scm}~\cite{scm}, which accompanies this paper, implements a general chain Binomial algorithm using the principles of Algorithm~\ref{al:cb}.
    This package contains a script, \texttt{examples/sir\_binomial.py}, which is used here to generate $500$ simulations of a stochastic SIR epidemic process over $400$ time steps with $\delta t = 0.25$.
    The initial conditions for the states are $(\tilde{S}(0), \tilde{I}(0), \tilde{R}(0)) = (990, 10, 0)$ and the transition rates are defined such that $\beta/\gamma=0.28/0.14=2$.
    Algorithm~\ref{al:r0}, also included in \texttt{scm}, is used to generate a distribution of $\tilde{\beta}/\tilde{\gamma}$ estimates i.e. $r_0$ estimates.
    An instance of running this code with these parameters is given in Figure~\ref{fig:Figure_1a_cb_ts}, it shows the timeseries of the states $\tilde{S}(t)$, $\tilde{I}(t)$ and $\tilde{R}(t)$ for each simulated process.
    Figures~\ref{fig:Figure_1b_cb_MeanPerSim} and \ref{fig:Figure_1c_cb_MeanAll} depict histograms of $\tilde{\beta}/\tilde{\gamma}$ estimates with quantiles $0.25$, $0.5$ and $0.75$ shown by the dashed lines.
    As expected the mode of the distribution of $\tilde{\beta}/\tilde{\gamma}$ estimates is very close to the model value $\beta/\gamma=2$.
    Given the mean $\overline{r}_0[j]$, i.e. the mean $\tilde{\beta}/\tilde{\gamma}$ from each simulation $j$, then the mean over all $500$ simulations is $2.025$ which is within $4$ standard errors ($n=500$) of the model value.
    This result is given in Table \ref{tab:binomialStepSize} along with other step sizes, all other simulations parameters are unchanged. 
    As can be seen reducing $\delta t$ brings the mean of the $\tilde{\beta}/\tilde{\gamma}$ distribution closer to the model value of $\beta/\gamma$, this is reasonable under a Euler–Maruyama approximation \cite{Gerald1994}.
    Hence when benchmarking this code/algorithm consideration should be given to the size of $\delta t$ since it influences the variability of $\tilde{\beta}/\tilde{\gamma}$ around the model value $\beta/\gamma$.

    \begin{table}[htb]
    \centering
    \begin{tabular}{l|l|l|l|l}
    $\delta t$ & time steps & mean     & SE     & number of SE from $\beta / \gamma$ \\ \hline
    1.0        & 100        & 2.153    & 0.00632  & 25                                   \\
    0.25       & 400        & 2.025    & 0.00736  & 4                                    \\
    0.025      & 4000       & 1.99796  & 0.00728  & 1                                         
    \end{tabular}
    \caption{Estimates of the mean reproduction number ($1/500 \sum_{500} \overline{r}_0[j]$) for a selection of time step sizes $\delta t$.  The standard error SE is computed with $n=500$ and the total simulation time period equals $\delta t \, \, \times$ (time steps) $= 100$.  As $\delta t$ decreases the mean approaches $\beta / \gamma = 2$.}
    \label{tab:binomialStepSize}
    \end{table}

    \subsection{Results for Gillespie algorithm}
\begin{figure}[htb]
        \centering
        \begin{subfigure}[t]{0.49\textwidth}
            \caption{Timeseries}
            \label{fig:Figure_2a_gl_ts}
            \vspace{-0.2cm}
            \centering
            \includegraphics[width=\linewidth]{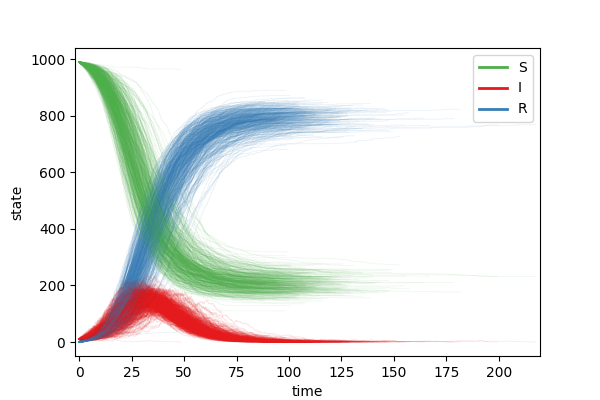} 
        \end{subfigure}

        \vspace{1.0cm}

        \begin{subfigure}[t]{0.49\textwidth}
            \caption{Histogram: mean $\tilde{\beta}/\tilde{\gamma}$}
            \label{fig:Figure_2b_gl_MeanPerSim}
            \vspace{-0.2cm}
            \centering
            \includegraphics[width=\linewidth]{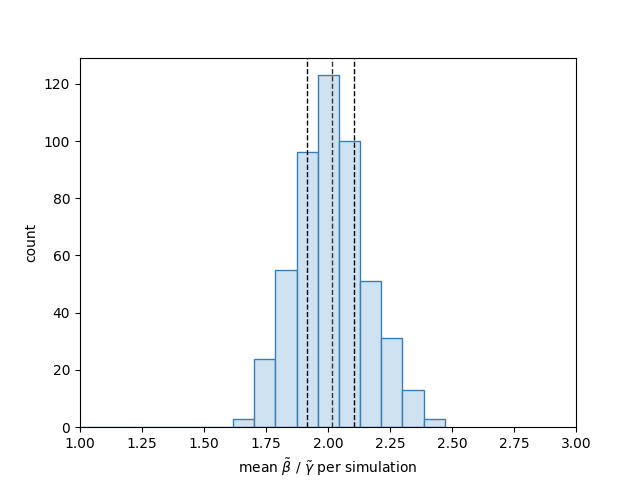} 
        \end{subfigure}
        \hfill
        \begin{subfigure}[t]{0.49\textwidth}
            \caption{Histogram: all $\tilde{\beta}/\tilde{\gamma}$}
            \label{fig:Figure_2c_gl_MeanAll}
            \vspace{-0.2cm}
            \centering
            \includegraphics[width=\linewidth]{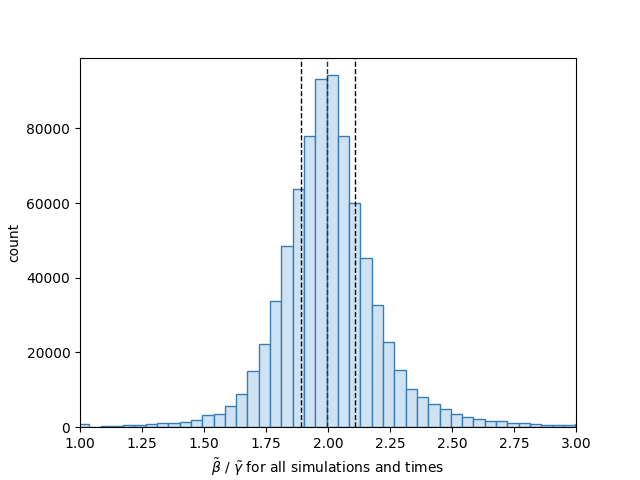} 
        \end{subfigure}

        \caption{500 simulations of the SIR epidemic process using the Gillespie algorithm.
        The top panel (\subref{fig:Figure_2a_gl_ts}) shows the timeseries for states $\tilde{S}(t)$, $\tilde{I}(t)$ and $\tilde{R}(t)$ for each simulated process.
        Panel (\subref{fig:Figure_2b_gl_MeanPerSim}) gives a histogram of the mean $\tilde{\beta}/\tilde{\gamma}$ from each simulation i.e. $\Bar{r}_0$.
        Lastly, panel (\subref{fig:Figure_2c_gl_MeanAll}) depicts a histogram of $\tilde{\beta}/\tilde{\gamma}$ at every time step, i.e. ${r}_0$, given all simulations.}
    \end{figure}

    A general Gillespie algorithm drawing on the principles of  Algorithm~\ref{al:gl} is included in the aforementioned package \texttt{scm}~\cite{scm}.
    Using this package $500$ simulations of a stochastic SIR epidemic process were computed with the script \texttt{examples/sir\_gillespie.py}. There were $1700$ time steps per simulation.
    As above $(\tilde{S}(0), \tilde{I}(0), \tilde{R}(0)) = (990, 10, 0)$ and $\beta/\gamma=0.28/0.14=2$.
    The timeseries for $\tilde{S}(t)$, $\tilde{I}(t)$ and $\tilde{R}(t)$ is given in Figure~\ref{fig:Figure_2a_gl_ts}, as expected the dynamics are very similar to those shown in Figure~\ref{fig:Figure_1a_cb_ts}.
    Figures~\ref{fig:Figure_2b_gl_MeanPerSim} and \ref{fig:Figure_2c_gl_MeanAll} depict histograms of $\tilde{\beta}/\tilde{\gamma}$ estimates with quantiles $0.25$, $0.5$ and $0.75$ shown by the dashed lines.
    The mode of the $\tilde{\beta}/\tilde{\gamma}$ estimates are, as expected, close to the model value $\beta/\gamma=2$.
    Given mean $\overline{r}_0[j]$, then the mean over all $j$ simulations was $2.013$ which is within $3$ standard errors of $\beta/\gamma=2$: note that in this instance $SE=0.0062$ with $n=500$.

    \subsection{Summary regarding SIR model tests}
    Given the results above it is concluded that when tested using Algorithm~\ref{al:r0} the computer code for the SIR model used in \texttt{scm}, and hence Algorithms~\ref{al:cb} and \ref{al:gl}, adhere to the underlying system of ODEs (Equations~\ref{eq:ODE1}~to~\ref{eq:ODE3}).
    Consequently it is confirmed that Algorithm~\ref{al:r0} could be used to both benchmark numerical results and write a unit test.
    For this model the distribution of $\tilde{\beta}/\tilde{\gamma}$ estimates are approximately Normal hence the mean can be used to estimate the mode.
    However if the same principles were applied to a different system of ODEs then Normality cannot be assumed.
    In addition to the aforementioned tests it would be prudent to check at every time step that 
    $\tilde{S}(t)+\tilde{I}(t)+\tilde{R}(t)=\mathcal{N}$ or equivalently $d\tilde{S}(t)/dt+d\tilde{I}(t)/dt+d\tilde{R}(t)/dt=0$.

\section{Lotka–Volterra model}
    The Lotka–Volterra, also called predator–prey, model \cite{Lotka1925, Volterra1926, Lotka1920, Volterra1928} is used as a second example to demonstrate how the general approach given in Section \ref{section:outlineMeth} can be applied beyond the SIR model.

    The Lotka–Volterra model may be used to describe the dynamics of a biological system where two species interact.
    One population consists of predators and the other of prey. 
    This model can be viewed as a graph having two nodes (compartments) that are connected by a bidirectional edge.
    The system of ODEs in terms of time $t$ is
    \begin{equation}
    \label{eq:ppODE1} 
    \frac{dx(t)}{dt} =  \alpha x(t) - \beta x(t)y(t),
    \end{equation}
    \begin{equation}
    \label{eq:ppODE2}
    \frac{dy(t)}{dt} =  \delta x(t)y(t) - \gamma y(t).
    \end{equation}
    The number of prey is denoted by $x(t)$ and number of predators by $y(t)$. Real constants $\alpha>0$, $\beta>0$, $\gamma>0$ and $\delta>0$ define the interaction between the two populations.
    The rate at which prey reproduce is represented by $\alpha x(t)$.
    The rate of predation upon these prey is proportional to the rate at which prey and predators meet, this is described by $\beta x(t)y(t)$.
    The rate of predator population growth is represented by $\delta x(t)y(t)$ and
    the loss rate of predators due to death or emigration is described by $\gamma y(t)$.
    
    It follows from Equation \ref{eq:ppODE1} and \ref{eq:ppODE2} that
    \begin{equation}
    \frac{dx(t)}{dy(t)} =  \frac{x(t)(\alpha - \beta y(t))}{y(t)(\delta x(t) - \gamma)}.
    \end{equation}
    Separation of variables may be used to rewrite this differential equation in integral form
    \begin{equation}
    \int_{x(0)}^{x(t)} \frac{\delta x - \gamma}{x} \,dx = \int_{y(0)}^{y(t)} \frac{\alpha - \beta y}{y} \,dy.
    \end{equation}
    Solving this integral equation and rearranging leads to
    \begin{equation}
    \label{eq:ppalpha}
    \alpha = \frac{\beta (y(t)-y(0)) + \delta(x(t)-x(0)) - \gamma \ln \left(x(t) \middle/ x(0) \right) }{\ln \left( y(t) \middle/ y(0) \right)  }.
    \end{equation}
    Note that Equation \ref{eq:ppalpha} in terms of $\gamma$ is:
    \begin{equation}
    \label{eq:ppgamma}
    \gamma = \frac{\beta (y(t)-y(0)) + \delta(x(t)-x(0)) - \alpha \ln \left( y(t) \middle/ y(0) \right) }{\ln \left( x(t) \middle/ x(0) \right)}.
    \end{equation}
    It follows that Equations \ref{eq:ppalpha} and \ref{eq:ppgamma} are not defined if the magnitude of a logarithm is infinity or either the numerator or denominator is zero.
    
    A predator-prey process can be simulated from Equations \ref{eq:ppODE1} and \ref{eq:ppODE2} using an algorithm suitable for stochastic compartmental models such as the chain Binomial or Gillespie algorithm.
    Given realisations from such a simulation the distribution of each parameter estimate is computed using Equation \ref{eq:ppalpha} and \ref{eq:ppgamma}: a given parameter estimate is equivalent to $r_0$ in Algorithm \ref{al:r0}.
    The mode of the distribution of these estimates is expected to equal its corresponding model parameter value.

\begin{figure}[htb]
        \centering
        \begin{subfigure}[t]{0.49\textwidth}
            \caption{Timeseries}
            \label{fig:Figure_3a_lv_ts}
            \vspace{-0.2cm}
            \centering
            \includegraphics[width=\linewidth]{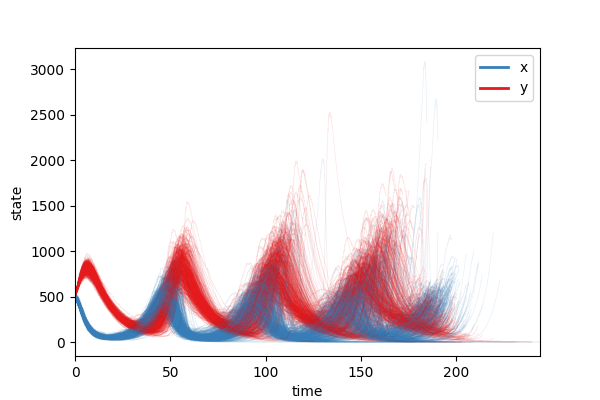} 
        \end{subfigure}

        \vspace{1.0cm}

        \begin{subfigure}[t]{0.49\textwidth}
            \caption{Histogram: $\alpha$ estimates}
            \label{fig:Figure_3b_lv_alpha}
            \vspace{-0.2cm}
            \centering
            \includegraphics[width=\linewidth]{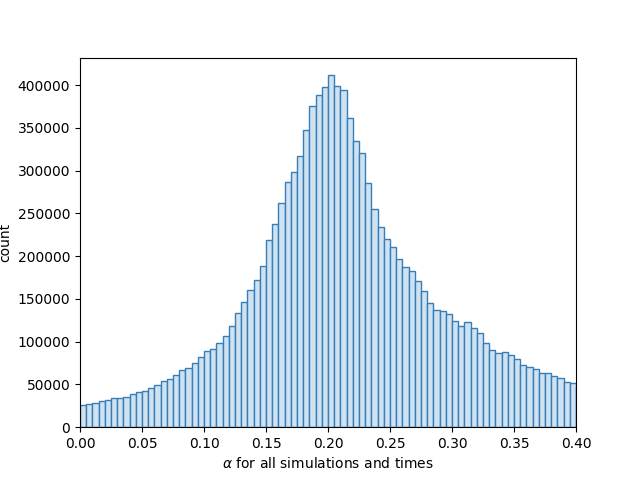} 
        \end{subfigure}
        \hfill
        \begin{subfigure}[t]{0.49\textwidth}
            \caption{Histogram: $\gamma$ estimates}
            \label{fig:Figure_3c_lv_gamma}
            \vspace{-0.2cm}
            \centering
            \includegraphics[width=\linewidth]{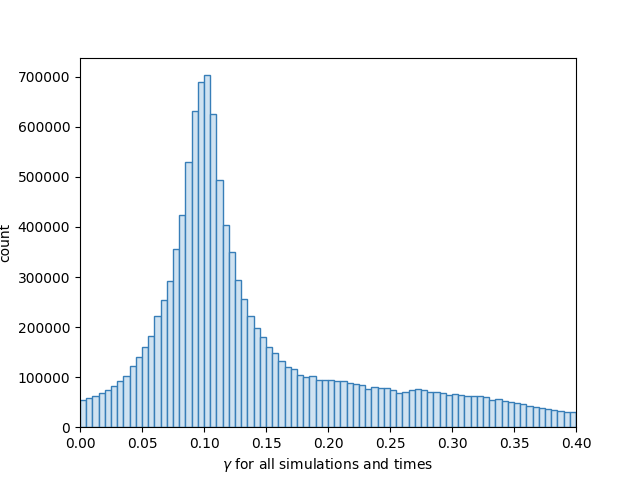} 
        \end{subfigure}

        \caption{500 simulations of the Lotka–Volterra biological process using the Gillespie algorithm.
        The top panel (\subref{fig:Figure_3a_lv_ts}) shows the prey $x(t)$ [blue] and predator $y(t)$ [red] timeseries for each simulated process.
        Panels (\subref{fig:Figure_3b_lv_alpha}) and (\subref{fig:Figure_3c_lv_gamma}) respectfully depict the histograms of the $\alpha$ and $\gamma$ parameter estimates over all times and simulations.}
    \end{figure}

    As an example let $(\alpha, \beta, \delta, \gamma)=(0.2, 0.0005, 0.0005, 0.1)$ and $(x(0), y(0))=(500, 500)$.
    With these parameters and initial conditions the solution to Equations \ref{eq:ppODE1} and \ref{eq:ppODE2} is a limit cycle oscillator. 
    Figure \ref{fig:Figure_3a_lv_ts} depicts the timeseries of $500$ simulations of the predator-prey process using the Gillespie algorithm: there are $30000$ time steps per simulation.
    Given these realisations then Equations \ref{eq:ppalpha} and \ref{eq:ppgamma} result in a distribution of estimates for $\alpha$ and $\gamma$ respectively.
    The distributions of these estimates are shown in Figures~\ref{fig:Figure_3b_lv_alpha} and \ref{fig:Figure_3c_lv_gamma}.
    As required Figure~\ref{fig:Figure_3b_lv_alpha} shows the mode of the distribution at  $\alpha \approx 0.2$ similarly the mode in Figure~\ref{fig:Figure_3c_lv_gamma} is at $\gamma \approx 0.1$.
    These figures show that the parameter estimate distributions are not Normal so the mode cannot be estimated from the mean.
    Nontrivial methods to estimate the mode of an arbitrary distribution are beyond the scope of this paper hence the use of histograms.

    The code used to simulate these results for the Lotka–Volterra process is available in \texttt{scm}~\cite{scm}, specifically see \texttt{examples/lotka\_volterra\_gillespie.py}.
    In this instance the core Gillespie algorithm code has already been tested using the SIR model therefore in practice this code does not necessarily need testing a second time using a different model e.g. the Lotka–Volterra model.
    Note that in this section the aim was to demonstrate that the techniques outlined above in relation to the SIR model are also applicable to other similar classes of ODE systems.

    In summary, the Lotka–Volterra model is considered in the context of stochastic compartmental model framework.
    In terms of the testing approach given in Section \ref{section:outlineMeth} this model differs from the SIR model in that the exact solution can be expressed in terms of any given parameter and the distribution of parameter estimates is not Normal.
    The computer code which estimated $x(t)$ and $y(t)$ has been shown to produce numerical results that are consistent with the Lotka–Volterra system of ODEs.
    This testing approach could therefore be used to benchmark numerical results, as such it could form the basis of a unit test.

\section{Discussion}
    In a real world setting it is likely that there will be discrepancies between the data collected from an actual epidemic or biological process, and the realisations of a stochastic compartmental model computed using either the chain Binomial or the Gillespie algorithm.
    In spite of this the computer code used to generate realisations from a stochastic compartmental model should not fail to adhere the underlying system of ODEs.
    It is of critical importance to test that such computer code generates plausible numerical output.
    Consequently it is vital that the numerical output is benchmarked in a controlled manner within a testing framework with robust test procedures such as those explored above.
    
    Software for stochastic compartmental models is often more general than the SIR model with three compartments given in Algorithm~\ref{al:cb} or \ref{al:gl}.
    For instance it may accommodate models with any number of compartments and the directed graph could include branching and feedback: it is straightforward to extend Algorithms~\ref{al:cb} and \ref{al:gl} to such cases.
    In this regard the computer code in \texttt{scm}~\cite{scm} for the chain Binomial and Gillespie algorithms respectively is quite general, although this particular implementation of the Gillespie algorithm does not include branching.
    As a consequence \texttt{scm} has the advantage of being applicable to any suitable compartmental model of arbitrary size and complexity.

    The main strength of the novel testing approach described in this paper is that it can be used to check that the numerical realisations adhere to the underlying system of ODEs independently of the algorithms used to generate these realisations.
    However this approach is limited in that depending on the system of ODEs it may not be possible to find solutions analytically in terms of every model parameter, or in terms of combinations of model parameters.
    By way of an example let the SIR model be extended such that it has an additional compartment $E$ between $S$ and $I$: this is the so-called SEIR model.
    To the author's knowledge it is not possible to solve this system of ODEs analytically in such a way that a solution can be written in terms of the transition parameter relating to $E \rightarrow I$.
    Specifically it is not possible to write an expression analogous to Equation \ref{eq:boverg} that involves the $E \rightarrow I$ transition parameter.
    However in terms of the underlying system of ODEs the first $S \rightarrow E$ and last $I \rightarrow R$ transition parameters of the SEIR model are equivalent to the $S \rightarrow I$ and last $I \rightarrow R$ parameters in the SIR model.
    Consequently one part of the SEIR model solution has the same form as Equation \ref{eq:boverg} therefore the novel testing approach developed in this paper could be directly applied.
    Any error in the code used to compute the realisations from an SEIR model would almost surely be exposed under such a test (using Equation \ref{eq:boverg}) due to the linear and sequential nature of the graph connecting the compartments.
    For stochastic compartmental models with more complicated graphs the testing methods explored in this paper may not be sufficient on their own or perhaps even applicable.
    
    In future work it would be of interest to explore the case where the system of ODEs for a stochastic compartmental model does not admit either a full or partial solution analytically. In this case it would be plausible to perform a test as follows:
    \begin{enumerate}
        \item compute the realisations of the process using for example the Gillespie algorithm;
        \item find the derivatives of these computed realisations for example by the finite difference method, see caveat below;
        \item rearrange one ODE in terms of a particular parameter and then compute estimates of that parameter using the aforementioned computed realisations and their derivatives;
        \item compute the mode of the resulting distribution of parameter estimates and check it matches the actual model 
        parameter value.
        \item repeat this procedure, where possible/required, for every parameter of interest in the system of ODEs.
    \end{enumerate}
    For example, although trivial, Equation \ref{eq:ODE1} could be written as $\beta = -\mathcal{N} I^{-1} S^{-1} \, dS/dt$, in which case substituting numerically computed realisations into the right side would result in a distribution of $\beta$ estimates where the mode of the distribution is the model parameter value of $\beta$.
    The advantage of this method is that it is more general than the approach explored in this paper since it does not rely on being able to integrate all, or a subset of, the system of ODEs analytically.
    However the caveat is that it relies on being able to compute derivatives of  realisations from a stochastic process, this may require a finite difference method with an order higher than first-order and/or a data smoothing method (e.g. weighted moving average) applied to the realisations prior to numerical differentiation.
    Although such a test has its place, it is also potentially problematic in that the logic is circular in so much as the numerical output (i.e. realisations of the process) from the code is directly fed back into the original system of ODEs.
    Hence, depending on how the test code is written, there is the possible danger that the numerical output (realisations), regardless of whether it is correct or erroneous, leads to its own confirmation!

    Often there is no completely foolproof way to test computer code.
    In the main the more general the code, the harder it is to thoroughly test.
    In terms of software which uses numerical methods it is prudent to benchmark the numerical output using a method which is as far removed as possible from the algorithm/code that was used to compute the original output.
    This was achieved in this paper by comparing solutions obtained analytically with numerical solutions.
    For general code designed for a class of models, e.g. stochastic compartmental model, `sufficiently' good test cases need to be designed.
    Here this could arguably be achieved by testing the software in \texttt{scm} using the SIR model as the test case.
    Ultimately during software development a judgement call will need to be made as to what constitutes a sufficient degree, and appropriate type, of testing.

    \bigskip    
    \bigskip    
  
    \textbf{Summary:} The novel key idea for benchmarking the solutions from stochastic compartmental models is to derive an exact, partial or full, solution analytically from the system of ODEs such that an expression can be written for the time dependent quantities in terms of a time independent quantity e.g. a model parameter.
    From this expression it then follows, given simulated realisations from an epidemic or biological process, that a distribution of time independent quantities is estimated.
    The mode of this distribution should equal the actual value of the time independent quantity.
    This procedure uses realisations from the entire simulation time interval without needing constraints such as the thermodynamic limit and/or long-time steady state limit.
    Furthermore these techniques could be applied to suitable systems of ODEs other than the SIR and Lotka–Volterra models.
    The novel techniques presented in this paper can therefore be used to create a robust test of the numerical solution produced by computer code used to generate realisations from stochastic compartmental models.

\section{Acknowledgements}
    The author would like to thank the Wellcome Trust for funding this research: grant `GEM: translational software for outbreak analysis'.

    This research was funded in whole, or in part, by the Wellcome Trust [Grant number]. For the purpose of Open Access, the author has applied a CC BY public copyright licence to any Author Accepted Manuscript version arising from this submission.

\bibliography{refs}{}

\begin{thebibliography}{10}

\bibitem{Tarlinder2016}
A.~Tarlinder, {\em Developer Testing: Building Quality into Software}.
\newblock The Addison-Wesley signature series, Addison-Wesley Professional,
  USA, 1st~ed., 2016.

\bibitem{Black2013}
R.~Black, {\em Pragmatic software testing: becoming an effective and efficient
  test professional}.
\newblock Wiley, Indianapolis, 1st~ed., 2013.

\bibitem{Myers2012}
G.~J. Myers, C.~Sandler, and T.~Badgett, {\em The art of software testing}.
\newblock Wiley, Hoboken, 3rd~ed., 2012.

\bibitem{Merali2010}
Z.~Merali, ``Computational science:...error ...why scientific programming does
  not compute,'' {\em Nature}, vol.~467, pp.~775--777, October 2010.

\bibitem{Fairley1978}
R.~E. Fairley, ``Tutorial: Static analysis and dynamic testing of computer
  software,'' {\em Computer}, vol.~11, no.~4, pp.~14--23, 1978.

\bibitem{Turek2006}
S.~Turek and J.~Hron, ``Proposal for numerical benchmarking of fluid-structure
  interaction between an elastic object and laminar incompressible flow,'' in
  {\em Fluid-Structure Interaction}, pp.~371--385, Springer, Berlin,
  Heidelberg, 2006.

\bibitem{Stadter2020}
P.~St{\"a}dter, Y.~Sch{\"a}lte, L.~Schmiester, J.~Hasenauer, and P.~L. Stapor,
  ``Benchmarking of numerical integration methods for ode models of biological
  systems,'' {\em Scientific Reports}, vol.~11, no.~2696, 2021.

\bibitem{Hickey2014}
J.~Hickey and J.~Gottsmann, ``Benchmarking and developing numerical finite
  element models of volcanic deformation,'' {\em Journal of Volcanology and
  Geothermal Research}, vol.~280, pp.~126--130, 2014.

\bibitem{lasson2018}
J.~R. de~Lasson, L.~H. Frandsen, P.~Gutsche, S.~Burger, O.~S. Kim,
  O.~Breinbjerg, A.~Ivinskaya, F.~Wang, O.~Sigmund, T.~H\"{a}yrynen, A.~V.
  Lavrinenko, J.~M{\o}rk, and N.~Gregersen, ``Benchmarking five numerical
  simulation techniques for computing resonance wavelengths and quality factors
  in photonic crystal membrane line defect cavities,'' {\em Optics Express},
  vol.~26, pp.~11366--11392, Apr 2018.

\bibitem{Abbey1952}
H.~Abbey, ``An examination of the reed-frost theory of epidemics,'' {\em Human
  biology}, vol.~24, no.~3, p.~201—233, 1952.

\bibitem{Fine1977}
P.~E.~M. Fine, ``{A commentary on the mechanical analogue to the Reed-Frost
  epidemic model},'' {\em American Journal of Epidemiology}, vol.~106, no.~2,
  pp.~87--100, 1977.

\bibitem{Gillespie1976}
D.~T. Gillespie, ``A general method for numerically simulating the stochastic
  time evolution of coupled chemical reactions,'' {\em Journal of Computational
  Physics}, vol.~22, no.~4, pp.~403--434, 1976.

\bibitem{Gillespie1977}
D.~T. Gillespie, ``Exact stochastic simulation of coupled chemical reactions,''
  {\em The Journal of Physical Chemistry}, vol.~81, no.~25, pp.~2340--2361,
  1977.

\bibitem{Brauer2019}
F.~Brauer, C.~Castillo-Chavez, and Z.~Feng, {\em Mathematical Models in
  Epidemiology}.
\newblock Springer-Verlag, New York, 1st~ed., 2019.

\bibitem{Lotka1925}
A.~J. Lotka, {\em Elements of physical biology}.
\newblock Williams \& Wilkins Company, Baltimore, 1925.

\bibitem{Volterra1926}
V.~Volterra, ``{Variazioni e fluttuazioni del numero d'individui in specie
  animali conviventi},'' {\em Memoria della Reale Accademia Nazionale dei
  Lincei}, vol.~2, pp.~31--113, 1926.

\bibitem{Kermack1927}
W.~O. Kermack and A.~G. McKendrick, ``A contribution to the mathematical theory
  of epidemics,'' {\em Proceedings of the Royal Society of London. Series A,
  Containing Papers of a Mathematical and Physical Character}, vol.~115,
  no.~772, pp.~700--721, 1927.

\bibitem{Hethcote2000}
H.~W. Hethcote, ``The mathematics of infectious diseases,'' {\em SIAM Review},
  vol.~42, no.~4, pp.~599--653, 2000.

\bibitem{Doob1942}
J.~L. Doob, ``Topics in the theory of markoff chains,'' {\em Transactions of
  the American Mathematical Society}, vol.~52, no.~1, pp.~37--64, 1942.

\bibitem{Doob1945}
J.~L. Doob, ``Markoff chains--denumerable case,'' {\em Transactions of the
  American Mathematical Society}, vol.~58, no.~3, pp.~455--473, 1945.

\bibitem{scm}
A.~C. Hale, {\em Stochastic Compartmental Models}, 2021.
\newblock Available at \url{https://gitlab.com/achale/scm/}.

\bibitem{Gerald1994}
C.~F. Gerald and P.~O. Wheatley, {\em Applied Numerical Analysis}.
\newblock Addison-Wesley Publishing Company, USA, 5th~ed., 1994.

\bibitem{Lotka1920}
A.~J. Lotka, ``Analytical note on certain rhythmic relations in organic
  systems,'' {\em Proceedings of the National Academy of Sciences}, vol.~6,
  no.~7, pp.~410--415, 1920.

\bibitem{Volterra1928}
V.~Volterra, ``{Variations and Fluctuations of the Number of Individuals in
  Animal Species living together},'' {\em ICES Journal of Marine Science},
  vol.~3, no.~1, pp.~3--51, 1928.

\end{thebibliography}
\bibliographystyle{ieeetr}

\bigskip

\end{document}